\documentstyle[preprint,aps]{revtex}
\tightenlines

\begin{document}
\title{Nonequilibrium thermodynamics of colloids}
\author{D. Bedeaux and J.M. Rubi*}
\address{Leiden Institute of Chemistry, University of Leiden, PO Box 9502, 2300 RA\\
Leiden, The Netherlands; *Departament de Fisica Fonamental, Universitat de\\
Barcelona, Diagonal 647, E-08028 Barcelona, Spain}
\date{\today }
\maketitle

\begin{abstract}
The common rules of classical nonequilibrium thermodynamics do not allow an
Onsager coefficient like the viscosity to depend on the shear rate. Such a
dependence is experimentally well documented, however. In this paper it is
shown, using nonequilibrium thermodynamics alone, how it is possible that
the effective viscosity coefficient of an isotropic colloidal system depends
on both the shear rate and the frequency.
\end{abstract}

\section{Introduction}

Much is known about the hydrodynamic description of colloidal systems. This
is described in a number of well written books on the subject [1,2,3]. In
these books a multitude of details about the interesting behavior of
suspended asymmetric colloidal particles in shear flow is treated. What is
not being done is to treat these systems in the context of nonequilibrium
thermodynamics. That such a treatment is possible becomes clear reading the
monograph by De Gennes [4,5] on The Physics of Liquid Crystals. Rather than
going into detail regarding the motion of individual particles he discusses,
both the equilibrium and the nonequilibrium properties of these materials,
using thermodynamics. In this paper we will apply nonequilibrium
thermodynamics to these colloidal systems. As it is not our intention to
duplicate all the well described knowledge about the collective behavior of
these systems, we will focus on one aspect which has remained puzzling from
the viewpoint of nonequilibrium thermodynamics.

It is well-known that the viscosity in these systems depends on the shear
rate. In view of the fact that the shear rate is not a thermodynamic state
variable, like, for instance, the temperature, classical nonequilibrium
thermodynamics, as explained in the classic monograph by de Groot and Mazur
[6], does not allow such a dependence. Books on colloidal hydrodynamics
explain this dependence in terms of the behavior of asymmetric suspended
particles in shear flow. The question we pose ourselves in this paper is:
explain this dependence using nonequilibrium thermodynamics alone.
Interestingly enough Prigogine and Mazur [7] already mentioned this problem
as one of the three possible examples of their method of internal variables.
The orientation of the suspended particles was suggested as such an internal
variable. In de Groot and Mazur's book [6] the other two examples are
treated in detail but this one was not pursued. It had to wait for a better
understanding of these systems. In this paper we will show that it is now
possible to answer this question for a suspension of asymmetric colloidal
particles. In addition to answering a question of principle, the treatment
is simpler and more direct than the current explanations. It is also not
restricted to dilute suspensions.

In the second section we give the Gibbs relation for an isotropic system
with asymmetric colloidal particles. The particles are assumed to have a
symmetry axis. A new term appears which gives the entropy change due to a
change in a tensorial order parameter ${\bf Q}$. This order parameter
characterizes the alignment of the particles [4,5]. This is precisely the
new internal variable Prigogine and Mazur [7] needed. In the third section
the resulting entropy production is presented. This leads to linear laws in
which the viscous pressure tensor and the time rate of change of the order
parameter are expressed in their conjugate thermodynamic forces. In the
fourth section these laws are solved for the special case of stationary
shear flow. An expression is found for the viscous pressure in terms of the
shear gradient. Their ratio defines an effective viscosity. This effective
viscosity is found to be a function of the shear. Crucial to obtain this
result is the fact that the Onsager coefficients in general depend on the
order parameter ${\bf Q}$ which is a state variable. This answers the
principle question. We proceed in chapter five with a analysis of time
dependent shear flow. It is found that this results in a frequency dependent
effective shear viscosity. In the last section we derive a simple expression
for the dependence of one of the Onsager coefficients on the order parameter
to make this point clear beyond doubt.

\section{The Gibbs relation}

In an isotropic suspension of particles, which are cylindrically symmetric,
a tensorial order parameter ${\bf Q}$ appears as a new thermodynamic
variables [4,5]. ${\bf Q}$ is a traceless symmetric tensor, which is defined
as the ensemble average of $\left( {\bf aa}-{\bf 1}/3\right) ,$ where ${\bf a%
}$ is a unit vector along the symmetry axis of one of the particles and $%
{\bf 1}$\ is the unit tensor. In an isotropic phase ${\bf Q}$\ is, in the
absence of any directing field, zero. In, for instance, a magnetic field $%
{\bf Q}$\ is nonzero. Changes of ${\bf Q}$\ lead to changes of the entropy.
It is therefore important to take the order parameter along in the
description of the system. The total differential of the entropy is given by
the following Gibbs relation 
\begin{equation}
Tds_{v}=du_{v}-\mu d\rho -\mu _{s}d\rho _{s}-{\bf w}:d{\bf Q}  \label{1.1}
\end{equation}
Here $T$\ is the temperature, $s_{v}=\rho s$\ the entropy density, $%
u_{v}=\rho u$ the internal energy density, $\mu $\ the chemical potential of
the solvent, $\rho $\ the solvent particle number density, $\mu _{s}$\ the
chemical potential of the suspended particles and $\rho _{s}$\ the suspended
particle number density. The densities with a subscript $v$, in addition to $%
\rho $\ and $\rho _{s},$ are per unit of volume. ${\bf w}$\ is the variable
conjugate to ${\bf Q}$\ and is given by 
\begin{equation}
{\bf w=-}T\left( \frac{\partial s_{v}}{\partial {\bf Q}}\right) _{u_{v},\rho
,\rho _{s}}  \label{1.2}
\end{equation}
Alternatively one may use the relation to the free energy $f_{v}$%
\begin{equation}
{\bf w=}\left( \frac{\partial f_{v}}{\partial {\bf Q}}\right) _{T,\rho ,\rho
_{s}}  \label{1.3}
\end{equation}
Furthermore $:$\ between two matrices signifies a double contraction. De
Gennes [4 (page 48),5] writes as explicit expression for the free energy
density 
\begin{equation}
f_{v}\left( T,\rho ,\rho _{s},{\bf Q}\right) =f_{v}^{0}\left( T,\rho ,\rho
_{s}\right) +\frac{1}{2}A\left( T,\rho ,\rho _{s}\right) tr\left( {\bf Q.Q}%
\right) {\bf +}\frac{1}{3}B\left( T,\rho ,\rho _{s}\right) tr\left( {\bf %
Q.Q.Q}\right) +\bigcirc \left( {\bf Q}^{4}\right)  \label{1.4}
\end{equation}
where $\bigcirc $\ is the order symbol. Here $tr$\ signifies the trace of a
matrix. Note that $tr\left( {\bf Q.Q}\right) ={\bf Q:Q}$. The resulting
traceless symmetric conjugate variable is 
\begin{equation}
{\bf w}=A\left( T,\rho ,\rho _{s}\right) {\bf Q+}B\left( T,\rho ,\rho
_{s}\right) \stackrel{\circ }{\overline{{\bf Q.Q}}}+\bigcirc \left( {\bf Q}%
^{3}\right)  \label{1.5}
\end{equation}
where $\stackrel{\circ }{\overline{{\bf Q.Q}}}$\ is the traceless symmetric
part of ${\bf Q.Q}$. This relation may be inverted to give the order
parameter in terms of its conjugate variable 
\begin{equation}
{\bf Q}=A^{-1}\left( T,\rho ,\rho _{s}\right) {\bf w-}A^{-3}\left( T,\rho
,\rho _{s}\right) B\left( T,\rho ,\rho _{s}\right) \stackrel{\circ }{%
\overline{{\bf w.w}}}+\bigcirc \left( {\bf w}^{3}\right)  \label{1.6}
\end{equation}
The typical size of $A$\ is of the order of $k_{B}T\rho _{s}$\ where $k_{B}$%
\ is Boltzmann's constant.

\section{The entropy production}

Without the ${\bf Q}$ term the Gibbs relation is equivalent to the one used
in de Groot and Mazur [6]. Note that they use densities per unit of mass
while we use densities per unit of volume like de Gennes [4,5] is doing. The
calculation of the entropy production, using the balance equations, is
essentially identical to the one they give. The only difference is the
appearance of a term due to the order parameter ${\bf Q}$. In order to
describe the contributions of the order parameter to the dynamical behavior
of the system we may restrict ourselves, using the Curie principle, to
contributions of force-flux pairs which are symmetric traceless tensors. The
rate of entropy production density, as a function of position ${\bf r}$\ and
time $t$,\ due to these contributions becomes 
\begin{equation}
T\sigma _{tens}\left( {\bf r},t\right) =-\stackrel{\circ }{\overline{{\bf %
\Pi }\left( {\bf r},t\right) }}:\stackrel{\circ }{\overline{grad{\bf v}%
\left( {\bf r},t\right) }}-{\bf w\left( r,t\right) :}\frac{d{\bf Q\left(
r,t\right) }}{dt}  \label{2.1}
\end{equation}
here $\stackrel{\circ }{\overline{{\bf \Pi }}}$\ is the symmetric traceless
part of the viscous pressure tensor and $\stackrel{\circ }{\overline{grad%
{\bf v}}}$\ the symmetric traceless part of the velocity gradient.

The resulting linear laws are 
\begin{eqnarray}
\stackrel{\circ }{\overline{{\bf \Pi }\left( {\bf r},t\right) }} &=&-2\eta 
\stackrel{\circ }{\overline{grad{\bf v}\left( {\bf r},t\right) }}-\ell {\bf %
w\left( r,t\right) }  \nonumber \\
\frac{d{\bf Q\left( r,t\right) }}{dt} &=&\ell \stackrel{\circ }{\overline{%
grad{\bf v}\left( {\bf r},t\right) }}-\ell _{Q}{\bf w\left( r,t\right) }
\label{2.2}
\end{eqnarray}
Here we used that, $\stackrel{\circ }{\overline{grad{\bf v}\left( {\bf r}%
,t\right) }}$ is odd and ${\bf w\left( r,t\right) }$\ even for time
reversal,so that the matrix of Onsager coefficients is antisymmetric. The
Onsager coefficients are in general functions of the thermodynamic state
variables, so that we have 
\begin{equation}
\eta \left( T,\rho ,\rho _{s},{\bf Q}\right) ,\qquad \ell \left( T,\rho
,\rho _{s},{\bf Q}\right) \quad \text{and}\quad \ell _{Q}\left( T,\rho ,\rho
_{s},{\bf Q}\right)  \label{2.3}
\end{equation}
In line with the assumption of linearity, they are not allowed to be
functions of $\stackrel{\circ }{\overline{grad{\bf v}\left( {\bf r},t\right) 
}}$\ and $d{\bf Q\left( r,t\right) /}dt$. The diagonal coefficient $\eta $\
is the viscosity and the diagonal coefficient $\ell _{Q}$\ is a typical
rotational mobility of the suspended particles divided by $\rho _{s}$. The
cross coefficient $\ell $\ is dimensionless and is expected to be a positive
constant of the order unity.

\section{Stationary elongational flow}

The above equations may easily be solved for stationary elongational flow 
\begin{equation}
{\bf v}\left( {\bf r},t\right) =\frac{1}{2}\stackrel{.}{\gamma }\left(
y,x,0\right)  \label{3.1}
\end{equation}
The resulting velocity gradient is 
\begin{equation}
\stackrel{\circ }{\overline{grad{\bf v}\left( {\bf r},t\right) }}=grad{\bf v}%
\left( {\bf r},t\right) =\frac{1}{2}\stackrel{.}{\gamma }\left( 
\begin{array}{ccc}
0 & 1 & 0 \\ 
1 & 0 & 0 \\ 
0 & 0 & 0
\end{array}
\right)  \label{3.2}
\end{equation}
Using the second linear law, eq.(\ref{2.2}b), results in 
\begin{equation}
{\bf w\left( r,t\right) }=\frac{\ell }{\ell _{Q}}\stackrel{\circ }{\overline{%
grad{\bf v}\left( {\bf r},t\right) }}=\frac{\ell }{2\ell _{Q}}\stackrel{.}{%
\gamma }\left( 
\begin{array}{ccc}
0 & 1 & 0 \\ 
1 & 0 & 0 \\ 
0 & 0 & 0
\end{array}
\right)  \label{3.3}
\end{equation}
Substitution in eq.(\ref{1.6}) gives 
\begin{equation}
{\bf Q}=\frac{\ell }{2A\ell _{Q}}\stackrel{.}{\gamma }\left( 
\begin{array}{ccc}
0 & 1 & 0 \\ 
1 & 0 & 0 \\ 
0 & 0 & 0
\end{array}
\right) +\bigcirc \left( \stackrel{.}{\gamma }^{2}\right)  \label{3.4}
\end{equation}
The first linear law, eq.(\ref{2.2}a), then yields 
\begin{equation}
\stackrel{\circ }{\overline{{\bf \Pi }\left( {\bf r},t\right) }}=-\left(
\eta +\frac{\ell ^{2}}{2\ell _{Q}}\right) \stackrel{.}{\gamma }\left( 
\begin{array}{ccc}
0 & 1 & 0 \\ 
1 & 0 & 0 \\ 
0 & 0 & 0
\end{array}
\right) \equiv -\eta _{eff}\stackrel{.}{\gamma }\left( 
\begin{array}{ccc}
0 & 1 & 0 \\ 
1 & 0 & 0 \\ 
0 & 0 & 0
\end{array}
\right)  \label{3.5}
\end{equation}
Writing the dependence on the thermodynamic state variables explicitly, the
effective viscosity becomes 
\begin{equation}
\eta _{eff}\left( T,\rho ,\rho _{s},{\bf Q}\right) =\eta \left( T,\rho ,\rho
_{s},{\bf Q}\right) +\frac{\ell ^{2}\left( T,\rho ,\rho _{s},{\bf Q}\right) 
}{2\ell _{Q}\left( T,\rho ,\rho _{s},{\bf Q}\right) }  \label{3.6}
\end{equation}
In view of the above relation of the order parameter with the rate of
elongation, one may write this relation as 
\begin{equation}
\eta _{eff}\left( T,\rho ,\rho _{s},\stackrel{.}{\gamma }\right) =\eta
\left( T,\rho ,\rho _{s},\stackrel{.}{\gamma }\right) +\frac{\ell ^{2}\left(
T,\rho ,\rho _{s},\stackrel{.}{\gamma }\right) }{2\ell _{Q}\left( T,\rho
,\rho _{s},\stackrel{.}{\gamma }\right) }  \label{3.7}
\end{equation}
The effective viscosity depends therefore on the rate of elongation.

\section{Oscillatory elongational flow}

The above equations may also easily be solved in the case of oscillatory
elongational flow. In that case the velocity field, its symmetric traceless
gradient, the order parameter and its conjugate variable all become
proportional to $\exp (-i\omega t)$. We now neglect the nonlinear terms in
eq.(\ref{1.5}) and the ${\bf Q}$ dependence\ of the transport coefficients
in order to avoid the appearance of higher order harmonics. Though this is
an interesting phenomenon, it is not our present concern. In the second
linear law we must now replace $d{\bf Q\left( r,t\right) /}dt$ by $-i\omega 
{\bf Q\left( r,t\right) }$. In the solution this implies that 
\begin{equation}
{\bf Q=}A^{-1}{\bf w}=\frac{\ell }{A\ell _{Q}-i\omega }\stackrel{\circ }{%
\overline{grad{\bf v}}}  \label{4.1}
\end{equation}
The resulting complex frequency dependent effective viscosity then becomes 
\begin{equation}
\eta _{eff}\left( T,\rho ,\rho _{s},\omega \right) =\eta \left( T,\rho ,\rho
_{s}\right) +\frac{\ell ^{2}\left( T,\rho ,\rho _{s}\right) }{2\ell
_{Q}\left( T,\rho ,\rho _{s}\right) \left( 1-i\omega \tau \left( T,\rho
,\rho _{s}\right) \right) }  \label{4.2}
\end{equation}
with a relaxation time 
\begin{equation}
\tau \left( T,\rho ,\rho _{s}\right) =\frac{1}{A\left( T,\rho ,\rho
_{s}\right) \ell _{Q}\left( T,\rho ,\rho _{s}\right) }  \label{4.3}
\end{equation}
This relaxation time is the reorientation time of the suspended particles
due to rotational diffusion, and is therefore of the order of the rotational
diffusion coefficient. It follows that $\omega \tau $\ is the relevant
P\'{e}clet number for rotational diffusion in an oscillating velocity field.
For low frequencies the effective viscosity is enhanced by the coupling to
the order parameter 
\begin{equation}
\eta _{eff}\left( T,\rho ,\rho _{s},\omega =0\right) =\eta \left( T,\rho
,\rho _{s}\right) +\frac{\ell ^{2}\left( T,\rho ,\rho _{s}\right) }{2\ell
_{Q}\left( T,\rho ,\rho _{s}\right) }  \label{4.4}
\end{equation}
This is the same result as found above, eq.(\ref{3.6}), for the stationary
case if one neglects the dependence of the coefficient on the order
parameter. For high frequencies one finds 
\begin{equation}
\eta _{eff}\left( T,\rho ,\rho _{s},\omega =\infty \right) =\eta \left(
T,\rho ,\rho _{s}\right)  \label{4.5}
\end{equation}
Writing the complex viscosity as the sum of a real and imaginary part, $\eta
_{eff}=\eta _{eff}^{\prime }+i\eta _{eff}^{\prime \prime }$, one has 
\begin{equation}
\eta _{eff}^{\prime }\left( T,\rho ,\rho _{s},\omega \right) =\eta \left(
T,\rho ,\rho _{s}\right) +\frac{\ell ^{2}\left( T,\rho ,\rho _{s}\right) }{%
2\ell _{Q}\left( T,\rho ,\rho _{s}\right) \left( 1+\omega ^{2}\tau
^{2}\left( T,\rho ,\rho _{s}\right) \right) }  \label{4.6}
\end{equation}
and 
\begin{equation}
\eta _{eff}^{\prime \prime }\left( T,\rho ,\rho _{s},\omega \right) =\frac{%
\omega \ell ^{2}\left( T,\rho ,\rho _{s}\right) }{2\ell _{Q}\left( T,\rho
,\rho _{s}\right) \left( 1+\omega ^{2}\tau ^{2}\left( T,\rho ,\rho
_{s}\right) \right) }  \label{4.7}
\end{equation}
This last expression gives the elastic contribution to the complex viscosity.

\section{The Onsager coefficients}

Nonequilibrium thermodynamics does not derive expressions for the Onsager
coefficients. Like the thermodynamic derivatives of the entropy as for
instance $A\left( T,\rho ,\rho _{s}\right) $, they have to be either
measured or calculated from a description in terms of the motion of the
separate particles. Above we have given estimates of the typical order of
magnitude of the various coefficients on the basis of the physical meaning
of the coefficients and dimensional arguments.

As the shear dependence of the effective viscosity is the property we set
out to find, we will pursue this element a bit further. One may write 
\begin{equation}
\eta =\eta _{0}\left( 1+\left[ \eta \right] \phi \right)  \label{5.1}
\end{equation}
where $\phi =V\rho _{s},$\ with $V$ the volume of the monodisperse suspended
particles. $\left[ \eta \right] $\ is the so-called intrinsic viscosity.
Each suspended particle contributes to the viscous dissipation. For low
densities the intrinsic viscosity is independent of their concentration.
Einstein [8] found that the intrinsic viscosity for spheres was equal to
2.5. For rods the contribution depends on the orientation ${\bf a}$\ of the
rod. In order to get the intrinsic viscosity one must then first calculate
the contribution of a rod to the viscous pressure tensor as a function of
its orientation and then average over the orientations [2].

The distribution over orientations may be calculated using the rotational
diffusion equation for this distribution. For the stationary state
distribution corresponding to a given order parameter ${\bf Q}$\ one has 
\begin{equation}
f\left( {\bf a}\right) =\frac{15}{8\pi }\left( {\bf aa-}\frac{{\bf 1}}{3}%
\right) :{\bf Q+}\frac{1}{4\pi }  \label{5.2}
\end{equation}

For spheroids, which are useful as a model system, the solution of the
velocity field has been given by Jeffery [9]. In the monograph by van de Ven
[2] the resulting contribution to the pressure tensor is given. In the
appendix we average those expressions over the above orientation
distribution. This results in 
\begin{eqnarray}
\left[ \eta \right] &=&\frac{2}{15}\left( r_{e}^{2}-1\right) \left[ \frac{%
5\ell }{4\left( 2r_{e}^{2}-\left( 1-2r_{e}^{2}\right) A_{e}\right) }\frac{%
\stackrel{.}{\gamma }}{A\ell _{Q}}\right.  \nonumber \\
&&\left. +\frac{26r_{e}^{2}-24r_{e}^{2}A_{e}-15A_{e}}{\left(
2r_{e}^{2}-3A_{e}\right) \left( 2r_{e}^{2}-\left( 2r_{e}^{2}+1\right)
A_{e}\right) }+\frac{6}{\left( r_{e}^{2}+1\right) \left( 3A_{e}-2\right) }%
\right]  \label{5.3}
\end{eqnarray}
Here the aspect ratio $r_{e}=a/b$\ is the ratio of the diameter $a$ along
the symmetry axis and the diameter $b$ normal to the symmetry axis.
Furthermore 
\begin{eqnarray}
A_{e} &=&\frac{r_{e}^{2}}{r_{e}^{2}-1}-\frac{r_{e}%
\mathop{\rm arccosh}%
r_{e}}{\left( r_{e}^{2}-1\right) ^{3/2}}\qquad \text{for }r_{e}>1  \nonumber
\\
A_{e} &=&\frac{r_{e}\arccos r_{e}}{\left( 1-r_{e}^{2}\right) ^{3/2}}-\frac{%
r_{e}^{2}}{1-r_{e}^{2}}\qquad \text{for }r_{e}<1  \label{5.4}
\end{eqnarray}
Note that $\stackrel{.}{\gamma }/A\ell _{Q}$ is the rotational P\'{e}clet\
number so that eq.(\ref{5.3}) is an expansion in the P\'{e}clet\ number. In
the limit of a sphere, $r_{e}\rightarrow 1,$ one may verify that the
expression reduces to 2.5. In that case the shear dependence of the
effective viscosity disappears. Prolate spheroids, $r_{e}>1$, are shear
thickening and oblate spheroids, $r_{e}<1$, are shear thinning for small
P\'{e}clet numbers in elongational flow [10].

\section{Discussion and conclusions}

The main objective was to show that a straightforward application of
nonequilibrium thermodynamics leads to a shear dependent viscosity. We were
able to derive a general nonlinear expression for this dependence. The
important reason for this is the coupling to the orientation of the
suspended colloidal particles. This orientation is described by a tensorial
order parameter ${\bf Q}$ as thermodynamic state variable. The fact that the
Onsager coefficients depend on this state variable then results in a shear
dependent viscosity. Of course the critical reader could have doubted such a
dependance of the Onsager coefficients on ${\bf Q.}$ In order to defend us
against this critique, we derive an explicit formula for small P\'{e}clet
numbers, $\stackrel{.}{\gamma }/A\ell _{Q}$. This gives a linear relation
for this dependence. For larger P\'{e}clet numbers the nonlinear relation
between ${\bf Q}$\ and ${\bf w}$\ must be taken into account. This will lead
to more realistic and interesting predictions. One interesting aspect is
that the powers of ${\bf Q}$\ lead to shear field with a different
''direction''. As the principle question, {\it why the viscosity depends on
the shear}, has been answered, we will not here pursue these points.

The frequency dependence is due to a single exponential decay of the order
parameter. For low frequencies the viscosity is enhanced by the coupling to
the orientation of the colloidal particles. For higher frequencies the
orientation lags behind. This leads to thinning of the fluid and a
reversible elastic response. One can extend the analysis, within the general
frame of nonequilibrium thermodynamics, by introducing the whole angular
distribution of the particles as an internal thermodynamic variable [6]. The
details for such an analysis have recently been worked out by Mazur [11].
This results in a whole spectrum of relaxation times and consequently gives
a more complex frequency dependence in the transition from low frequency to
high frequency behavior.

In our analysis we do not need to analyse the orientational distribution
function in any detail. The knowledge of the order parameter is enough. This
makes our analysis much simpler that the usual procedure, where the
rotational diffusion equation for this distribution function is solved
numerically. As we said above [11], the orientational distribution function
can be introduced as an internal thermodynamic variable. In addition to a
richer frequency dependence this will also lead to a more complex shear
dependence.

The low frequency and the low shear behavior of the viscosity has a very
similar non-analytic dependance on these parameters [12]. This is a general
phenomenon which is not specifically due to the asymmetry, or for that
matter the presence, of the suspended particles. It falls outside the scope
of the present article.

In the field of socalled extended irreversible thermodynamics [13], one
assumes that fluxes, fluxes of fluxes, etc., are also thermodynamic state
variables. This assumption distinguishes it from classical irreversible
thermodynamics, where this is not considered to be a proper assumption. The
viscous pressure is such a flux and is, according to extended irreversible
thermodynamics, a state variable. On the basis of this assumption the
viscosity can be a function of the viscous pressure. As a result, the
viscosity then becomes a function of the shear. The assumption, so to say,
defines the problem away. We will not try to invalidate their assumption. We
only want to stress that the present paper shows that such an assumption is
not needed to obtain a shear dependent viscosity. As such their assumption
seems to be one to many.

{\large Acknowledgement}

We want to thank T.G.M. van de Ven for a clarification of some of the
formulae in his book. D. Bedeaux wants to thank DGICYT\ for supporting a
visit to the University of Barcelona where this work was started. J.M. Rubi
wants to thank N.W.O. for support of a visit to Leiden University where part
of this work was finalized.

\section{Appendix}

We use the results of Jeffery [9] as presented in van de Ven's monograph
[2], see in particular pages 229-231. The shear field he uses is $%
v_{1}=v_{2}=0$ and $v_{3}=Gx_{2}$. In our notation $x=x_{2},y=x_{3},z=x_{1},%
\phi =c$ and $\stackrel{.}{\gamma }=G$. The symmetric traceless part of this
simple shear field is identical to the one we used above. In view of the
linearity of the problem we may therefore use the symmetric traceless part
of the contribution to the viscous pressure tensor van de Ven uses: 
\begin{equation}
{\bf \Pi =\Pi }_{0}+\frac{8\pi \eta _{0}}{V}\phi \left\langle {\bf M}%
\right\rangle  \label{A.1}
\end{equation}
where $V=4\pi b^{3}r_{e}/3$\ is the volume of the particle. ${\bf \Pi }%
_{0}=-2\eta _{0}grad{\bf v}\left( {\bf r},t\right) $ is the viscous pressure
one would have in the absense of the suspended particles. He gives the
tensor ${\bf M}$ as ${\bf M}^{\prime }$ in eq.(3.280) in a coordinate frame
attached to the particle. Using the distribution given above, eq.(\ref{5.2}%
), we have 
\begin{equation}
\left\langle {\bf M}\right\rangle =\frac{15}{8\pi }\left[ \int {\bf M}\left( 
{\bf a}\right) \left( {\bf aa-}\frac{{\bf 1}}{3}\right) d\Omega \right] :%
{\bf Q}+\frac{1}{4\pi }\int {\bf M}\left( {\bf a}\right) d\Omega  \label{A.2}
\end{equation}
where $d\Omega =\sin \theta d\theta d\phi $ with $0<\theta <\pi $\ and $%
0<\phi <2\pi $. The integral between square brackets gives a constant times
the four index symmetric traceless unit tensor ${\bf \Delta }$. The matrix
elements of this tensor are given by 
\begin{equation}
\Delta _{ijkl}=\frac{1}{2}\delta _{ik}\delta _{jl}+\frac{1}{2}\delta
_{il}\delta _{jk}-\frac{1}{3}\delta _{ij}\delta _{kl}  \label{A.3}
\end{equation}
The constant can be calculated by taking $i=l$, $j=k$\ and summing over $i$\
and $j$. This results in 
\begin{equation}
\left\langle {\bf M}\right\rangle =\frac{3}{16\pi }\left[ \int {\bf M}\left( 
{\bf a}\right) :\left( {\bf aa-}\frac{{\bf 1}}{3}\right) d\Omega \right] 
{\bf Q}+\frac{1}{4\pi }\int {\bf M}\left( {\bf a}\right) d\Omega  \label{A.4}
\end{equation}
The contraction of two matrices is independent of the reference frame and we
may therefore use the reference frame attached to the particle. In this
reference frame $\left( {\bf aa-1/}3\right) =\left( 2/3\right) {\bf e}_{1}%
{\bf e}_{1}-\left( 1/3\right) {\bf e}_{2}{\bf e}_{2}-\left( 1/3\right) {\bf e%
}_{3}{\bf e}_{3}$ in terms of the unit vectors in this frame. This gives 
\begin{equation}
\left\langle {\bf M}\right\rangle =\frac{1}{16\pi }\left[ \int \left(
2M_{11}^{\prime }-M_{22}^{\prime }\left( \theta ,\phi \right)
-M_{33}^{\prime }\left( \theta ,\phi \right) \right) d\Omega \right] {\bf Q}+%
\frac{1}{4\pi }\int {\bf M}\left( \theta ,\phi \right) d\Omega  \label{A.5}
\end{equation}
Using eq.(3.280) we see that $M_{22}^{\prime }$\ and $M_{33}^{\prime }$\
integrate to zero while $M_{11}^{\prime }$\ is independent of direction. In
the second integral we can use eqs.(3.282)-(3.284) to obtain the only
nonzero contribution. The result is 
\begin{equation}
\left\langle {\bf M}\right\rangle =\frac{1}{2}M_{11}^{\prime }{\bf Q}+\frac{%
b^{3}r_{e}}{6}\left( \frac{4}{15}C_{1}+\frac{1}{3}C_{2}+\frac{2}{3}%
C_{3}\right) \stackrel{.}{\gamma }\left( 
\begin{array}{ccc}
0 & 1 & 0 \\ 
1 & 0 & 0 \\ 
0 & 0 & 0
\end{array}
\right)  \label{A.6}
\end{equation}
Substitution of ${\bf Q}$ then yields 
\begin{equation}
\left\langle {\bf M}\right\rangle =\left[ \frac{\ell }{4A\ell _{Q}}%
M_{11}^{\prime }+\frac{b^{3}r_{e}}{6}\left( \frac{4}{15}C_{1}+\frac{1}{3}%
C_{2}+\frac{2}{3}C_{3}\right) \right] \stackrel{.}{\gamma }\left( 
\begin{array}{ccc}
0 & 1 & 0 \\ 
1 & 0 & 0 \\ 
0 & 0 & 0
\end{array}
\right)  \label{A.7}
\end{equation}
Substitution into eq.(\ref{A.1}) and comparing with eq.(\ref{5.1}) gives for
the intrinsic viscosity 
\begin{equation}
\left[ \eta \right] =\left[ \frac{\ell }{A\ell _{Q}}\frac{3M_{11}^{\prime }}{%
2b^{3}r_{e}}+\frac{4}{15}C_{1}+\frac{1}{3}C_{2}+\frac{2}{3}C_{3}\right]
\label{A.8}
\end{equation}
Substitution of eqs.(3.280) and (3.284) finally gives eq.(\ref{5.3}).

\end{document}